\documentclass[fleqn,10pt]{article}
\usepackage{abstract}
\usepackage{amsmath}
\usepackage{amsopn}
\usepackage{amssymb}
\usepackage{authblk}
\usepackage{comment}
\usepackage[margin=0.75in]{geometry}
\usepackage{graphicx}
\usepackage{hyperref}
\usepackage{epstopdf}
\usepackage{color}
\usepackage{titlesec}
\usepackage[super,sort&compress]{natbib}


\title{{\bf Navigability of Random Geometric Graphs in the Universe and Other Spacetimes}}

\author[1]{William Cunningham}
\author[2]{Konstantin Zuev}
\author[3]{Dmitri Krioukov}

\affil[1]{\small {\it Department of Physics, Northeastern University, 360 Huntington Ave., Boston, MA 02115, United States}}
\affil[2]{{\it Department of Computing and Mathematical Sciences, California Institute of Technology, 1200 E.\ California Blvd., Pasadena, CA 91125, United States}}
\affil[3]{{\it Department of Physics, Department of Mathematics, Department of Electrical \& Computer Engineering, Northeastern University, 360 Huntington Ave., Boston, MA 02115, United States}}
\date{}

\newcommand{\munu}{{\mu\nu}}
\newcommand{\dprime}{{\prime\prime}}
\DeclareMathOperator{\arctanh}{arctanh}

\renewcommand{\thesection}{\Roman{section}}
\titleformat{\section}[block]{\Large\bfseries\filcenter}{\thesection.\!}{12pt}{}
\renewcommand{\thesubsection}{\Alph{subsection}}
\titleformat{\subsection}[block]{\large\bfseries\filcenter}{\thesubsection.\!}{12pt}{}

\begin{document}
\maketitle

\vspace{-0.5cm}
\begin{abstract}
Random geometric graphs in hyperbolic spaces explain many common structural and dynamical properties of real networks, yet they fail to predict the correct values of the exponents of power-law degree distributions observed in real networks. In that respect, random geometric graphs in asymptotically de Sitter spacetimes, such as the Lorentzian spacetime of our accelerating universe, are more attractive as their predictions are more consistent with observations in real networks. Yet another important property of hyperbolic graphs is their navigability, and it remains unclear if de Sitter graphs are as navigable as hyperbolic ones. Here we study the navigability of random geometric graphs in three Lorentzian manifolds corresponding to universes filled only with dark energy (de Sitter spacetime), only with matter, and with a mixture of dark energy and matter. We find these graphs are navigable only in the manifolds with dark energy. This result implies that, in terms of navigability, random geometric graphs in asymptotically de Sitter spacetimes are as good as random hyperbolic graphs. It also establishes a connection between the presence of dark energy and navigability of the discretized causal structure of spacetime, which provides a basis for a different approach to the dark energy problem in cosmology.
\end{abstract}

\section{Introduction}
Random geometric graphs~\cite{Dall2002Random,Penrose03-book,Spodarev2013StochasticGeometry} formalize the notion of ``discretization'' of a continuous geometric space or manifold. Nodes in these graphs are points, sprinkled randomly at constant sprinkling density, over the manifold, thus representing ``atoms'' of space, while links encode geometry---two nodes are connected if they happen to lie close in the space. These graphs are also a central object in algebraic topology since their clique complexes~\cite{Costa2015Fundamental} are Rips complexes~\cite{Hausmann1995VietorisRips,Kahle2011RGC} whose topology is known to converge to the manifold topology under very mild assumptions~\cite{Latschev2001VietorisRips}.

In network science and applied mathematics, random geometric graphs have attracted increasing attention over recent years~\cite{Serrano2012Uncovering,Kleineberg2016Hidden,Allard2017Geometric,Bianconi2015Perspective,Ostilli2015Statistical,Bianconi2015ComplexEvolution,Wu2015Emergent,Bianconi2016Network,Bianconi2017Emergent,Zhang2014Opinion,Newman2015Generalized,Henderson2011Geometric,Roberts2016Contribution,Javarone2013Perception,Xie2015Modeling,Xie2015Random,Jin2017Coupling,Clough2016What,Clough2016Embedding,Asta2015Geometric,Gugelmann2012Random,Fountoulakis2015Geometrization,Bode2015Largest,Candellero2016Bootstrap,Candellero2016Clustering,Abdullah2015Typical,Fountoulakis2016Law,Bringmann2015Geometric,Bringmann2016Average,Bringmann2016Greedy,Bradonjic2010Efficient,Bubeck2014Testing,Dhara2016Solvable,friedrich2015cliques,friedrich2015diameter,blasius2016hyperbolic,Penrose2013SRGG}, since it was shown that if the space defining these graphs is not Euclidean but negatively curved, i.e., hyperbolic, then these graphs provide a geometric explanation of many common structural and dynamical properties of many real networks, including scale-free degree distributions, strong clustering, community structure, and network growth dynamics~\cite{KrPa10,PaBoKr11,Zuev2015GPA}. Yet more interestingly, these graphs also explain the optimality of many network functions related to finding paths in the network without global knowledge of the network structure~\cite{BoKrKc08,BoKr09}. Random hyperbolic graphs appear to be optimal, that is, maximally efficient, with respect to the greedy path finding strategy that uses only spatial geometry to navigate through a complex network structure by moving at each step from a current node to its neighbor closest to the destination in the space~\cite{KrPa10,Bringmann2016Greedy}. The efficiency of this process is called network navigability~\cite{kleinberg00-nature}. High navigability of random hyperbolic graphs led to practically viable applications, including the design of efficient routing in the future Internet~\cite{BoPa10,Lehman2016Hyperbolic}, and demonstration that the spatiostructural organization of the human brain is nearly as needed for optimal information routing between different parts of the brain~\cite{Gulyas2015Navigable}. Yet if random hyperbolic graphs are truly geometric, meaning that if the sprinkling density is indeed constant with respect to the hyperbolic volume form, then the exponent $\gamma$ of the distribution $P(k)\sim k^{-\gamma}$ of node degrees $k$ in the resulting graphs is exactly $\gamma=3$~\cite{KrPa10}. In contrast, in random geometric graphs in de Sitter spacetime, which is asymptotically the spacetime of our accelerating universe, or indeed in the spacetime representing the exact large-scale Lorentzian geometry of our universe, this exponent asymptotically approaches $\gamma=2$~\cite{KrKi12}, as in many real networks~\cite{BoLaMoChHw06}. Yet it remains unclear if these random Lorentzian graphs are as navigable as random hyperbolic graphs.

In physics, random geometric graphs in Lorentzian spacetimes, directed in the time direction,
are known as causal sets, a central object in the causal set approach to quantum gravity~\cite{ref:bombelli1987}. A seemingly unrelated big, if not the biggest unsolved problem in cosmology is the dark energy puzzle~\cite{Albrecht2006}. What is dark energy? Why is its density orders of magnitude smaller than one would expect from high-energy physics? Causal sets provide one of the simplest explanation attempts to date~\cite{So07}, but there are many other attempts~\cite{Weinberg1987,GaLi99,ArHa00,Bousso2006,Bousso2011Geometric,BaSh11,HaSh12}, none commonly considered to be the final answer.

Here we study the navigability of undirected random geometric graphs in three Lorentzian manifolds. One manifold is de Sitter spacetime, corresponding to a universe filled with dark energy only, and no matter. Another manifold is the other extreme, a universe filled only with dust matter, and no dark energy. The third manifold is a universe like ours, containing both matter and dark energy. This last manifold interpolates between the other two. At early times and small graph sizes, it is matter-dominated and ``looks'' like the dust-only spacetime. At later times and large graph sizes, it is dark-energy-dominated and ``looks'' increasingly more like de Sitter spacetime.

We find that random geometric graphs only in manifolds with dark energy are navigable. Specifically, if there is no dark energy, that is, in the dust-only spacetime, there is a finite fraction of paths for which geometric path finding fails, and that this fraction is constant---it does not depend on the cutoff time, i.e., the present cosmological time in the universe, if the average degree in the graph is kept constant. In contrast, in spacetimes with dark energy, i.e., de Sitter spacetime and the spacetime of our universe, the fraction of unsuccessful paths quickly approaches zero as the cutoff time increases.

For network science this finding implies that in terms of navigability, random geometric graphs in Lorentzian spacetimes with dark energy are as good as random hyperbolic graphs. For physics, this finding establishes a connection between the presence of dark energy and navigability of the discretized causal structure of spacetime, which provides a basis for a different approach to the dark energy problem.

\subsection{Lorentzian Manifolds}
While Riemannian manifolds are manifolds with positive-definite metric tensors $g_{ij}$ defining geodesic distances $ds$ by $ds^2=\sum_{i,j=1}^dg_{ij}\,dx_i\,dx_j$, where $d$ is the manifold dimension, Lorentzian manifolds are manifolds whose metric tensors $g_\munu$, $\mu,\nu=\{0,1,\ldots,d\}$, have signature $(-++\ldots+)$, meaning that if diagonalized by a proper choice of the coordinate system, these tensors have one negative entry on the diagonal, while all other entries are positive. In general relativity, Lorentzian manifolds represent relativistic spacetimes, which are solutions of Einstein's equations. Typically, the dimension of a Lorentzian manifold is denoted by $d+1$, with the ``+1'' referring to the temporal (zeroth) dimension, while the other $d$ dimensions are spatial. In this paper we consider only $(3+1)$-dimensional Lorentzian manifolds, that is, manifolds of dimension equal to the dimension of our universe~\cite{Tegmark1997Dimensionality}. The Lorentzian metric structure naturally defines spacetime's causal structure: timelike intervals with $\Delta s^2<0$ connect pairs of causally related events, i.e., timelike-separated points on a manifold. \par

Einstein's equations are a set of ten coupled non-linear partial differential equations:
\begin{equation}
\label{eq:einstein}
R_\munu - \frac12 Rg_\munu + \Lambda g_\munu = 8\pi T_\munu\,,
\end{equation}
where we use the natural units with the gravitational constant and speed of light set to unity. The Ricci curvature tensor $R_\munu$ and Ricci scalar $R$ measure the manifold curvature, the cosmological constant $\Lambda$ is proportional to the dark energy density in the spacetime, and the stress-energy tensor $T_\munu$ represents the matter content. Spacetimes which are homogeneous and isotropic are called Friedmann-Robertson-Walker (FRW) spacetimes, which have a metric of the form $ds^2 = -dt^2 + a(t)^2d\Sigma^2$. The time-dependent function $a(t)$ in front of the spatial metric $d\Sigma$ is called the scale factor. This function characterizes the expansion of the volume form in a spatial hypersurface with respect to time; it alone tells whether there is a ``Big Bang'' at $t=0$, i.e., whether $a(0)=0$. The scale factor is derived explicitly as a solution to the $00$-component ($\mu=\nu=0$) of~\eqref{eq:einstein}, known as the first Friedmann equation:
\begin{equation}
\label{eq:friedmann}
\left(\frac{\dot a}{a}\right)^2 = \frac \Lambda 3 - \frac K a^2 + \frac{c}{a^{3g}}\,.
\end{equation}
The variable $g$ represents the type of matter in the spacetime: in this work we use the values $g=\{0,1\}$ to indicate no matter and dust matter, respectively. The spatial curvature of the spacetime is captured by $K$: $K=\{+1,0,-1\}$ implies positive, zero, or negative spatial curvature, respectively. Motivated by the observation that our universe is nearly flat~\cite{ref:komatsu2011}, in this work we use $K=0$, which significantly simplifies the calculations below. In the flat case, the spatial metric, hereafter using dimensionless spherical coordinates, becomes $d\Sigma^2 = dr^2 + r^2\,d\theta^2 + r^2\sin^2\theta\,d\phi^2$. Finally, $c$ is a constant proportional to the density of matter in the universe. \par

The total energy density in the universe is known to come from four sources: the matter (dark and baryonic) density $\rho_M$, the dark energy density $\rho_\Lambda$, the radiation energy density $\rho_R$, and the curvature $K$. The densities may be rescaled by a critical density: $\Omega\equiv\rho/\rho_c$, where $\rho_c\equiv3H_0^2/8\pi$; $H_0\equiv\dot{a}_0/a_0$ is the Hubble constant and $a_0\equiv a(t_0)$, i.e., the scale factor at the present time. Similarly, the curvature density parameter may be written as $\Omega_K\equiv-K/(a_0H_0)^2$ so that we obtain the state equation $\Omega_M + \Omega_\Lambda + \Omega_R + \Omega_K = 1$. This allows us to rewrite~\eqref{eq:friedmann} in the integral form~\cite{ref:weinberg2008}
\begin{equation}
\label{eq:friedmann2}
H_0t = \int_0^{a/a_0}\!\frac{dx}{x\sqrt{\Omega_\Lambda + \Omega_Kx^{-2} + \Omega_Mx^{-3} + \Omega_Rx^{-4}}}\,.
\end{equation}

In the flat universe, the curvature energy density contribution is zero: $\Omega_K=0$. Furthermore, except for a short period in the early universe, the radiation energy density is also negligible compared to the other terms: $\Omega_R\approx 0$. Therefore, we study manifolds defined only by $\Omega_\Lambda$ and $\Omega_M$: the de Sitter (dark energy only) manifold ($\Lambda > 0, g=c=0$), the Einstein-de Sitter (dust only) manifold ($\Lambda = 0, g=1, c>0$), and the mixed dark energy and dust manifold ($\Lambda,c > 0, g=1$). Hereafter, these three manifolds are respectively referred to as the energy (\emph{E}), dust (\emph{D}), and mixed (\emph{M}) manifolds. Defining rescaled time $\tau=t/\lambda$, the scale factors in these spacetimes are solutions to~\eqref{eq:friedmann2}, respectively using non-zero $\Omega_\Lambda$, $\Omega_M$, or both:
\begin{equation}
\label{eq:scale_factors}
a_E(\tau) = \lambda e^{\tau}\,, \quad
a_D(\tau) = \alpha\left(\frac{3}{2}\tau\right)^{2/3}\,, \quad
a_M(\tau) = \alpha\sinh^{2/3}\left(\frac{3}{2}\tau\right)\,.
\end{equation}
The parameters $\lambda$ and $\alpha$ respectively define the temporal and spatial scales. In a de Sitter manifold, there is no distinction between temporal and spatial scales, so that there is no $\alpha$, because the generators of the Lorentz group $SO(1,3)$ form a proper subset of those of the de Sitter group $SO(1,4)$, thereby removing a degree of freedom in the model. In manifolds which represent spacetimes with dust matter, this symmetry is broken, and relative rescalings between $\lambda$ and $\alpha$ are equivalent to an isotropic rescaling of space with respect to time. \par

The spatial scale of a mixed manifold, such as the one approximating our real universe, arises naturally from~\eqref{eq:friedmann2} when dimensionless variables are used; it is defined as $\alpha\equiv a_M(t_0)(\Omega_M/\Omega_\Lambda)^{1/3}$, related to the relative amount of dark energy~\cite{KrKi12}. The scale factor $a_M(\tau)$ asymptotically matches $a_D(\tau)$ at earlier times (a hot, matter-dominated universe) and $a_E(\tau)$ at later times (a cold, dark energy-dominated universe), so that the mixed manifold can be characterized by the dark energy density parameter $\Omega_\Lambda$. This way, the dark energy density is a measure of time via $\tau = (2/3)\arctanh\sqrt{\Omega_\Lambda}$. Using the present-day value of $\Omega_{\Lambda,0}\approx0.737$ in our universe gives the current rescaled cosmological time $\tau_0=t_0/\lambda\approx 0.473$, so that $\lambda$ sets the spacetime's timescale~\cite{ref:astier2006}. \par

In the FRW spacetimes defined by~\eqref{eq:scale_factors}, the scale factor and the metric tensor are used to find the volume form of the manifold:
\begin{equation}
\label{eq:diff_volume}
dV = \sqrt{-|g_\munu|}\sin\theta\,dt\,dr\,d\theta\,d\phi = a(t)^3r^2\sin\theta\,dt\,dr\,d\theta\,d\phi\,,
\end{equation}
where $r$ is the dimensionless radial coordinate and $\theta$ and $\phi$ are the polar and azimuthal angular coordinates. To study a particular spacetime in simulations below, it is necessary to consider its compact region, bounded by a temporal cutoff $t\in[0,t_0]$ and radial cutoff $r\in[0,r_0]$. Using rescaled temporal and spatial cutoffs $\tau_0=t_0/\lambda$ and $\rho_0=\tilde\alpha r_0$, where $\tilde\alpha=\alpha/\lambda$, except de Sitter spacetime where $\rho_0=r_0$, the volume of such a region in each spacetime is easily obtained via the integration of~\eqref{eq:diff_volume} within the corresponding bounds:
\begin{equation}
\label{eq:volumes}
V_E\left(\tau_0,\rho_0\right) = \frac{4\pi}{9}\lambda^4 \rho_0^3\left(e^{3\tau_0}-1\right)\,, \quad
V_D\left(\tau_0,\rho_0\right) = \pi\lambda^4\rho_0^3\tau_0^3\,, \quad
V_M\left(\tau_0,\rho_0\right) = \frac{2\pi}{9}\lambda^4\rho_0^3\left(\sinh\left(3\tau_0\right) - 3\tau_0\right)\,.
\end{equation}
\par

We will also use conformal time $\eta$, defined as $\eta(t) = \int^tdt^\prime/a(t^\prime)$, which is
\begin{equation}
\eta_E\left(\tau\right) = -e^{-\tau}\,, \quad
\eta_D\left(\tau\right) = \frac{1}{\tilde\alpha}\left(12\tau\right)^{1/3}\,, \quad
\eta_M\left(\tau\right) = \frac{2}{\tilde\alpha}\sinh^{1/3}\left(\frac{3}{2}\tau\right){}_2F_1\left(\frac{1}{6},\frac{1}{2};\frac{7}{6};-\sinh^2\left(\frac{3}{2}\tau\right)\right)\,,
\end{equation}
where ${}_2F_1$ is the hypergeometric function. This transformation is particularly useful for distinguishing between timelike and spacelike intervals, since in these coordinates, the scale factor may be factored out: $ds^2 = a^2(t(\eta))(-d\eta^2+d\Sigma^2)$, so that timelike and spacelike intervals with $\Delta s^2<0$ and $\Delta s^2>0$ correspond to intervals with $\Delta\eta^2>\Delta\Sigma^2$ and $\Delta\eta^2<\Delta\Sigma^2$, respectively.

\begin{figure}[!t]
\includegraphics[width=0.5\textwidth]{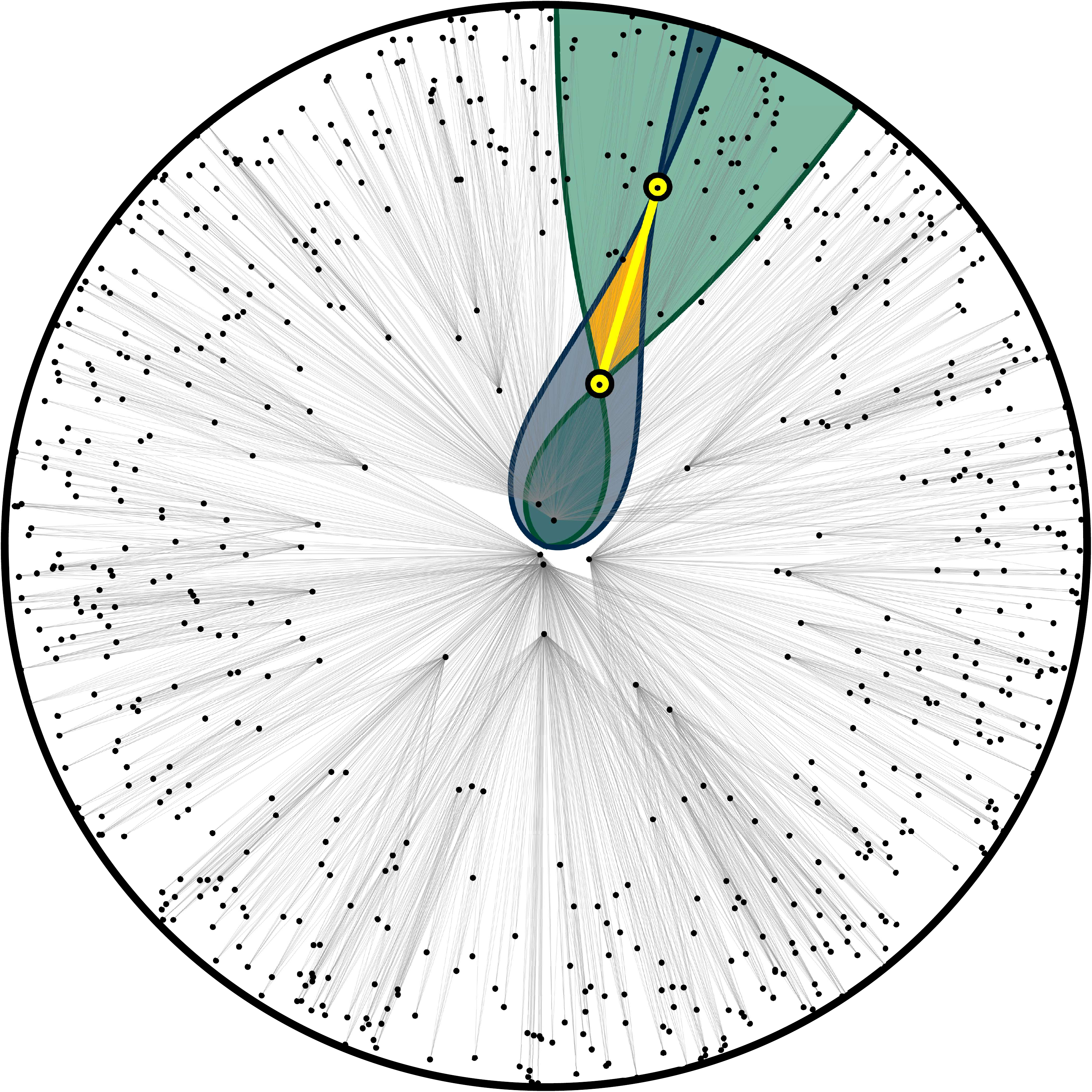}
\centering
\caption{\textbf{Random geometric graph in (1+1)-dimensional de Sitter spacetime.} The graph is realized by Poisson sprinkling $700$ nodes onto a $(1+1)$-dimensional de Sitter manifold, with compact spatial foliation by circles, which are hypersurfaces of constant time. The temporal cutoff is $\tau_0=5.94$, which is the radius of the disk shown. In the figure, the graph has been mapped from the de Sitter manifold to a disk of this radius by equating the time coordinates of all points in de Sitter spacetime with the radial coordinates in the shown disk. A pair of nodes, shown in yellow, is chosen and their light cones are shown in gray and green. The yellow nodes are connected to all other nodes that happen to lie in their corresponding light cones. In particular, the yellow nodes are connected to each other since they lie within each other's light cones. The overlap between the past and future light cones of the higher-$t$ and lower-$t$ yellow nodes respectively, shown in orange, is their Alexandroff set. The full set of grey links is obtained by iterating over all node pairs.}
\label{fig1}
\end{figure}

\subsection{Random Geometric Graphs}
Given a compact region of any $d$-dimensional manifold $\mathcal{M}$, a geometric graph $G_{\mathcal{M}}(N,R)$ in it is a set of $N$ nodes $n=\{n_1,\ldots,n_N\}$ with coordinates $x=\{x_1,\ldots,x_N\}$, and undirected edges connecting pairs $(n_i,n_j)$ located at distance $D(x_i,x_j)<R$ in the manifold~\cite{Penrose03-book}. Such a graph is called a random geometric graph (RGG) when the coordinates $x_i$ are a realization of a Poisson or other random point process, thereby defining an ensemble of RGGs. Directed Lorentzian RGGs $G_{\mathcal{M}}(n,0)$, also known as causal sets~\cite{ref:bombelli1987}, ``converge'' to Lorentzian manifolds $\mathcal{M}$ in the thermodynamic limit $N\to\infty$, since the causal structure alone is enough to recover the topology of a Lorentzian manifold~\cite{ref:hawking1976,ref:malament1977}. While the simplest base (open sets) of the manifold topology in the Riemannian case are open balls, this base in the Lorentzian case are Alexandroff sets, which are intersections of past and future light cones of points in the manifold~\cite{ref:alexandrov1937,ref:kronheimer1967}. Therefore, an undirected Lorentzian RGG is constructed by Poisson sprinkling points onto $\mathcal{M}$, and then linking those pairs which are timelike separated, hence $R=0$. \par

\section{Results}
\subsection{Constructing Random Geometric Graphs in Lorentzian Manifolds}
We construct RGGs in Lorentzian manifolds by sampling three spatial and one temporal coordinates for $N$ nodes in a particular region using a Poisson point process: $N$ is a random variable sampled from the Poisson distribution with mean $\bar{N}$, giving a sprinkling density $\delta\equiv N/V$. Given volumes~\eqref{eq:volumes}, and using the rescaled sprinkling density $q=\delta\lambda^4$, the numbers of nodes in the three spacetimes are given by
\begin{equation}
\label{eq:nodes}
N_E\left(\tau_0,\rho_0\right) = \frac{4\pi}{9} q \rho_0^3\left(e^{3\tau_0}-1\right)\,, \quad
N_D\left(\tau_0,\rho_0\right) = \pi q \rho_0^3\tau_0^3\,, \quad
N_M\left(\tau_0,\rho_0\right) = \frac{2\pi}{9} q \rho_0^3\left(\sinh\left(3\tau_0\right) - 3\tau_0\right)\,,
\end{equation}
\begin{figure}[!t]
\includegraphics[width=\textwidth]{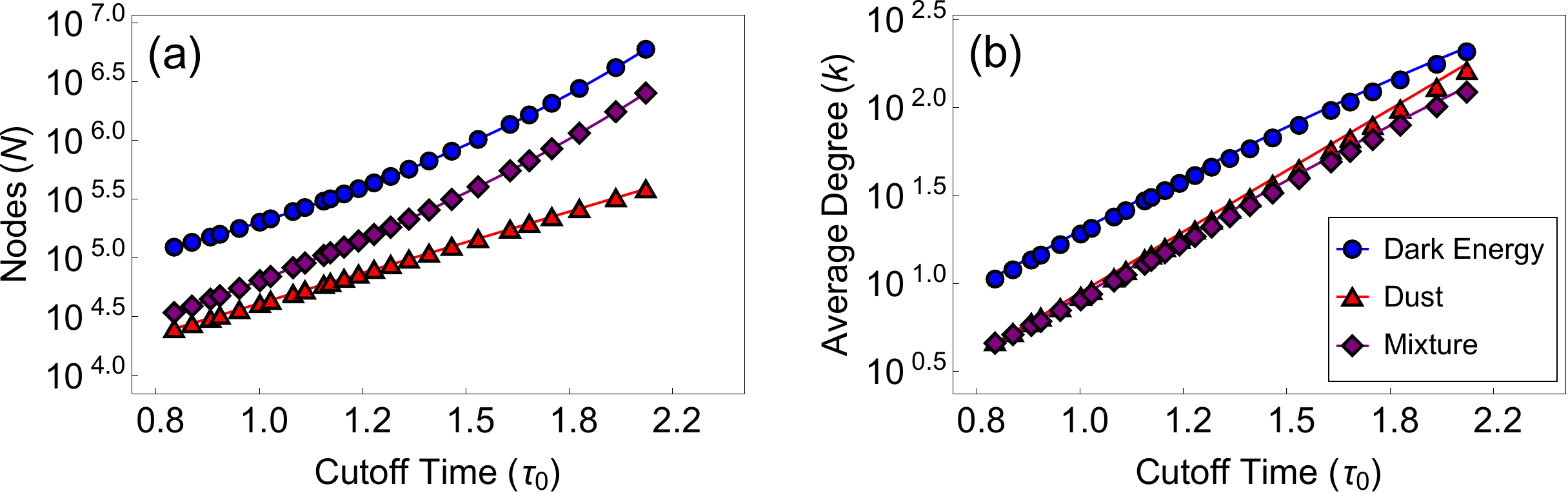}
\centering
\caption{\textbf{Graph size and average degree as functions of the cutoff time.} The figure shows the graph size $N$ and average degree $\bar{k}$ in simulations versus theoretical predictions, the solid curves, given by~(\ref{eq:nodes},\ref{eq:degrees}), for the constant rescaled sprinkling density $q=60$ and spatial cutoff $\rho_0=6$.}
\label{fig3}
\end{figure}
where all the parameters $q,\rho_0,\tau_0$ are dimensionless. The distributions $P(t)$, $P(r)$, $P(\theta)$, and $P(\phi)$ used to sample the coordinates are found in~(\ref{eq:scale_factors},~\ref{eq:diff_volume}) by comparing it to the volume form $dV = \chi(t)\chi(r)\chi(\theta)\chi(\phi)\,dt\,dr\,d\theta\,d\phi$, where $\chi$'s are the non-normalized density functions, $\chi(t)=a(t)^3$, $\chi(r)=r^2$, $\chi(\theta)=\sin\theta$, $\chi(\phi)=1$, and $\chi()\propto P()$ upon normalization, i.e., $P(t)=\chi(t)/\int_0^{t_0}\chi(t')\,dt'$, for instance. A pair of nodes $(n_i,n_j)$ is timelike related and, therefore, linked in the resulting graph if the following inequality is true:
\begin{equation}
\label{eq:spatial_form}
\Delta\Sigma_{ij}^2 = r_i^2 + r_j^2 - 2r_ir_j\left(\cos\theta_i\cos\theta_j + \sin\theta_i\sin\theta_j\cos\left(\phi_i - \phi_j\right)\right) < \left(\eta_i - \eta_j\right)^2\,,
\end{equation}
where the law of cosines has been used for the spatial distance $\Delta\Sigma_{ij}$ between the two nodes in three dimensions. Figure~\ref{fig1} visualizes a random geometric graph in $(1+1)$-dimensional de Sitter spacetime, where $\Delta\Sigma_{ij}^2=(\theta_i-\theta_j)^2$ instead of~\eqref{eq:spatial_form}. \par

In simulations in the next section, we will also need to generate graphs with a given average degree. To find the expected average degree in RGGs in our Lorentzian regions, we observe that the volume of the past and future light cones emanating from any given node, and bounding regions timelike-related to the node, is directly proportional, with the proportionality coefficient $1/\delta$, to the expected number of sprinkled nodes in them, and consequently, to the expected past and future degrees of the node. Integrating the expressions for these volumes, weighted by the node density in the space, over the entire region provides a theoretical expression for the expected degree as a function of the rescaled sprinkling density $q=\delta\lambda^4$ and the rescaled temporal cutoff $\tau_0=t_0/\lambda$, we get:
\begin{align}
\label{eq:degrees}
\begin{aligned}
\bar{k}_E(\tau_0) &= \frac{4\pi q}{9}\frac{\left(e^{-\tau_0}-1\right)\left(13-e^{-\tau_0}\left(14-13e^{-\tau_0}\right)\right)+6\tau_0\left(e^{-3\tau_0}+1\right)}{1-e^{-3\tau_0}}\,, \\
\bar{k}_D(\tau_0) &= \frac{18\pi q}{385}\tau_0^4\,, \\
\bar{k}_M(\tau_0) &= \frac{8\pi q}{\sinh\left(3\tau_0\right) - 3\tau_0}\int_0^{\tau_0}\!d\tau^\prime\int_0^{\tau_0}\!d\tau^\dprime\,\sinh^2\left(\frac{3\tau^\prime}{2}\right)\sinh^2\left(\frac{3\tau^\dprime}{2}\right)|\tilde{\eta}_M\left(\tau^\prime\right) - \tilde{\eta}_M\left(\tau^\dprime\right)|^3\,,
\end{aligned}
\end{align}
where rescaled conformal time $\tilde{\eta}\equiv\tilde\alpha\eta$ is used for convenience. These expressions do not depend on spatial cutoff $\rho_0$ because they are approximations for spatially large regions with $\rho_0\gg\tau_0$, so that boundary effects, i.e., the contributions to the average degree from nodes with $\rho$s close to $\rho_0$, are negligible. \par

It is evident from the exposition above including (\ref{eq:nodes},~\ref{eq:degrees}) that the only three out of the original five parameters defining the RGG ensemble with $N$ nodes and average degree $\bar{k}$---sprinkling density $\delta\equiv N/V$, temporal scale $\lambda$, spatial scale $\alpha$, and temporal and spatial cutoffs $t_0$ and $r_0$---are independent because $N$ depends only three dimensionless parameters, $q$, $\rho_0$, and $\tau_0$, while $\bar{k}$ depends only on two, $q$ and $\tau_0$. This is because sprinkling density $\delta$ sets the discreteness scale, which can be rescaled by $\lambda$: two graph ensembles with different $\delta$s and $\lambda$s are the same if their rescaled sprinkling density $q=\delta\lambda^4$ is the same. Similarly, two graph ensembles with different $\lambda$s and $t_0$s are the same, if their $\tau_0$s are the same, and two graph ensembles, and even spacetime regions, with different $\alpha$s and $r_0$s are the same, if their $\rho_0$s are the same. Therefore parameters $q,\rho_0,\tau_0$ form one natural choice of independent parameters, the one we use in simulations below. Yet any three independent functions of these parameters is an equivalent choice. In particular, $N$ and $\bar{k}$ are two such independent functions, so that $N,\bar{k},\tau_0$ is another choice of parameters that we also use in simulations. Yet we note that one parameter in these two sets of three parameters is not entirely independent, because the spatial cutoff $\rho_0$ must be such that $\rho_0\gg\tau_0$, so that the spatial boundary effects are negligible, and approximations~\eqref{eq:degrees} are valid, see Figure~\ref{fig3} and Methods. \par

\begin{figure}[!t]
\includegraphics[width=\textwidth]{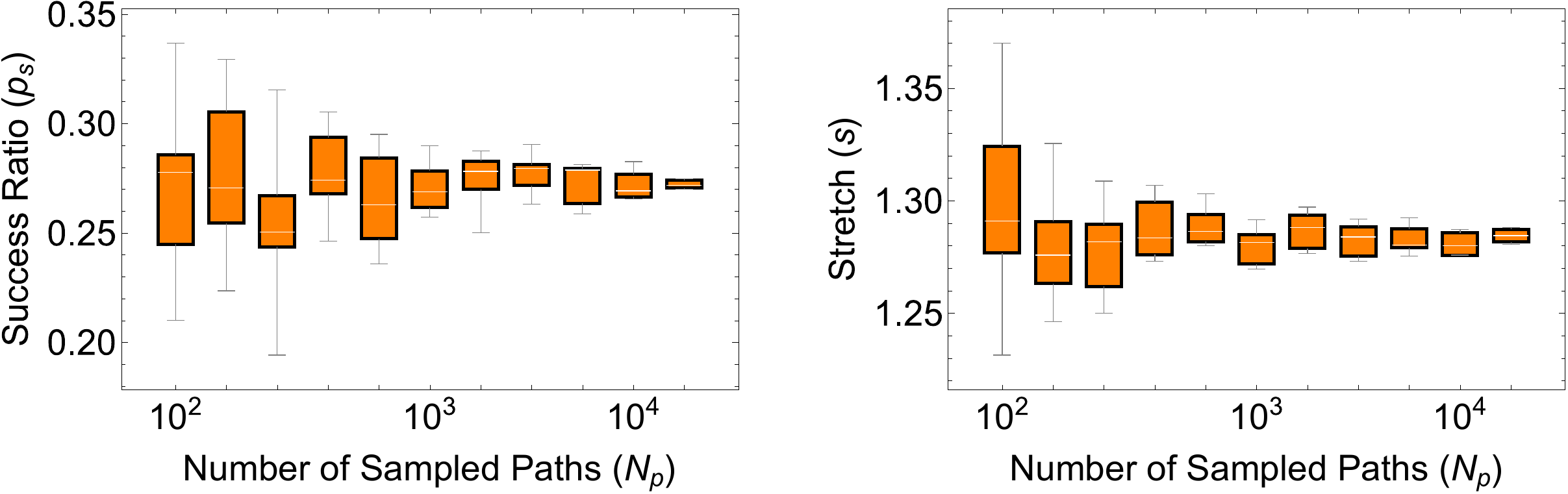}
\centering
\caption{\textbf{Convergence of success ratio and stretch.} The boxplots summarize the distributions of the success ratio (a) and stretch (b) as functions of the number $N_p$ of random source-destination node pairs sampled in $10$ random geometric graphs ($N_p$ pair samples in each graph) in the Einstein-de Sitter (dust) manifold with $\tau_0=4.64$, $\bar{k}=10$, and $N=2^{20}$. The orange boxes range from the first to third quartiles, while the bars are minima and maxima. The distributions stabilize at $N_p\ll N$.}
\label{figS1}
\end{figure}

\subsection{Navigability of Random Geometric Graphs in Lorentzian Manifolds}
The navigability of a geometric graph is the efficiency of greedy geometric path finding in it. This path finding strategy uses only local nearest-neighbor information to find a path in the graph between a given source node and a given destination node. Starting with the source node, the next node on the path is determined as the node's neighbor closest to the destination node according to geodesic distances in the manifold. When the closest neighbor has already been visited, the greedy path enters a loop. It does not reach the destination and is thus unsuccessful. This situation occurs when the two nodes forming the loop, also called a local minimum, do not have any third node that would be closer to the destination than the two nodes. The success ratio $p_s$ is defined as the fraction of greedy paths which successfully reach their destination, across a given set of source-destination node pairs in the graph. Here we select $N$ such pairs uniformly at random, where $N$ is the graph size. Increasing the number of pairs above $N$ does not noticeably affect the results, as can be seen from Figure~\ref{figS1}.
Another navigability metric is the stretch. The stretch of a successful greedy path is the ratio of the length of the path, measured as the number of hops, to the length of the shortest path between the same source and destination in the graph. The average stretch is the average of this quantity across successful paths between a given set of source-destination node pairs. \par

The geodesic distance between a pair of nodes on the underlying manifold is found by integrating the geodesic differential equations: $\frac{\partial^2x^\mu}{\partial\sigma^2} + \Gamma_{\nu\rho}^\mu\frac{\partial x^\nu}{\partial\sigma}\frac{\partial x^\rho}{\partial\sigma} = 0$, with $\mu,\nu,\rho=\{0,1,2,3\}$, where $x^\mu(\sigma) = (t(\sigma), r(\sigma), \theta(\sigma), \phi(\sigma))$ is a geodesic curve in four dimensions parameterized by an affine parameter $\sigma$, and $\Gamma^\mu_{\rho\nu} = \frac{1}{2}g^{\mu\beta}\left(\frac{\partial g_{\beta\nu}}{\partial x^\rho}+\frac{\partial g_{\beta\rho}}{\partial x^\nu}-\frac{\partial g_{\nu\rho}}{\partial x^\beta}\right)$, with $\beta=\{0,1,2,3\}$, are the Christoffel symbols which specify the affine connections in a curved space. The convention of summation over repeating indices is assumed. For a pair of nodes with temporal coordinates $t_1$ and $t_2$ and spatial distance $\Delta\Sigma_{1,2}$, the integration of this equation yields the geodesic distance:
\begin{equation}
d_{1,2} = \int_{t_1}^{t_2}\!\sqrt{\left|\frac{-\psi a^2\left(t\right)}{1+\psi a^2\left(t\right)}\right|}\,dt\,, \quad
\Delta\Sigma_{1,2} = \int_{t_1}^{t_2}\!\left(a^2\left(t\right)+\psi a^4\left(t\right)\right)^{-1/2}\,dt\,,
\end{equation}
where the parameter $\psi$ is found by solving the second transcendental equation numerically. \par

\begin{figure}[!pt]
\includegraphics[width=\textwidth]{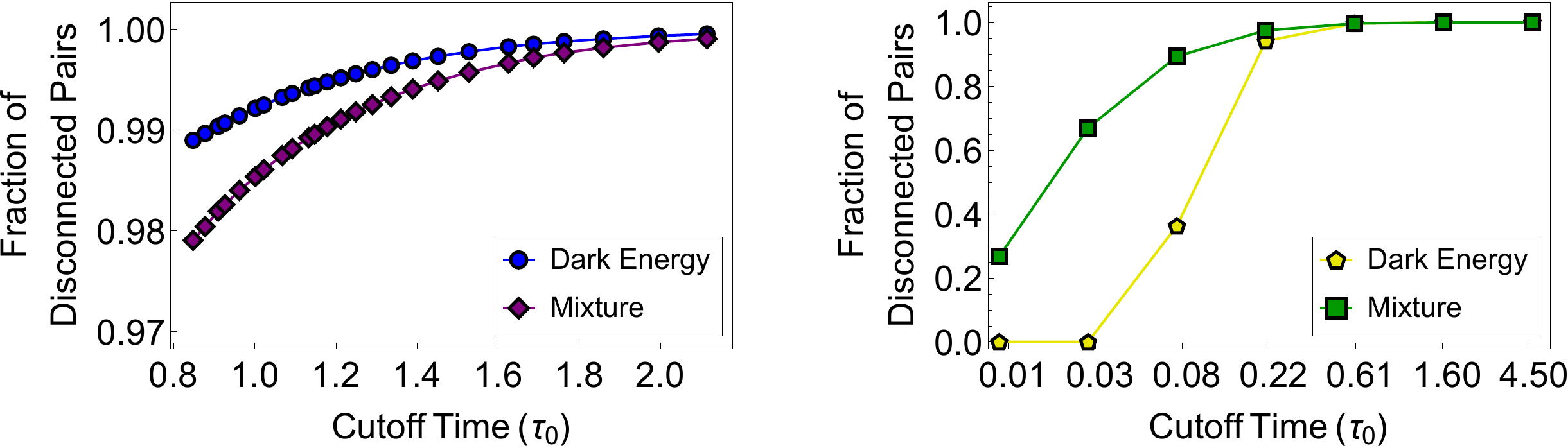}
\centering
\caption{\textbf{Fraction of geodesically disconnected node pairs} in the graphs in Figs.~2-4~in the main text. Panels~(a,b) correspond to the graphs in the de Sitter (dark energy) and mixed manifolds with $q=60, \rho_0=6$ and $N=2^{20}, \bar{k}=10$, respectively. The graphs in the Einstein-de Sitter (dust) manifold have trivially no geodesically disconnected node pairs since the manifold is geodesically connected.}
\label{figS2}
\end{figure}

As opposed to Riemannian manifolds, Lorentzian manifolds can be geodesically incomplete, i.e., there can exist pairs of spacelike separated points between which a geodesic does not exist~\cite{ref:oneill1983}. For such geodesically disconnected source-destination pairs, geodesic distances and consequently geodesic routing are \emph{undefined}, so that we exclude such pairs from our calculations. The percentages of geodesically disconnected node pairs in random graphs in the experiments below are reported in Figure~\ref{figS2}. \par

\begin{figure}[!pb]
\includegraphics[width=\textwidth]{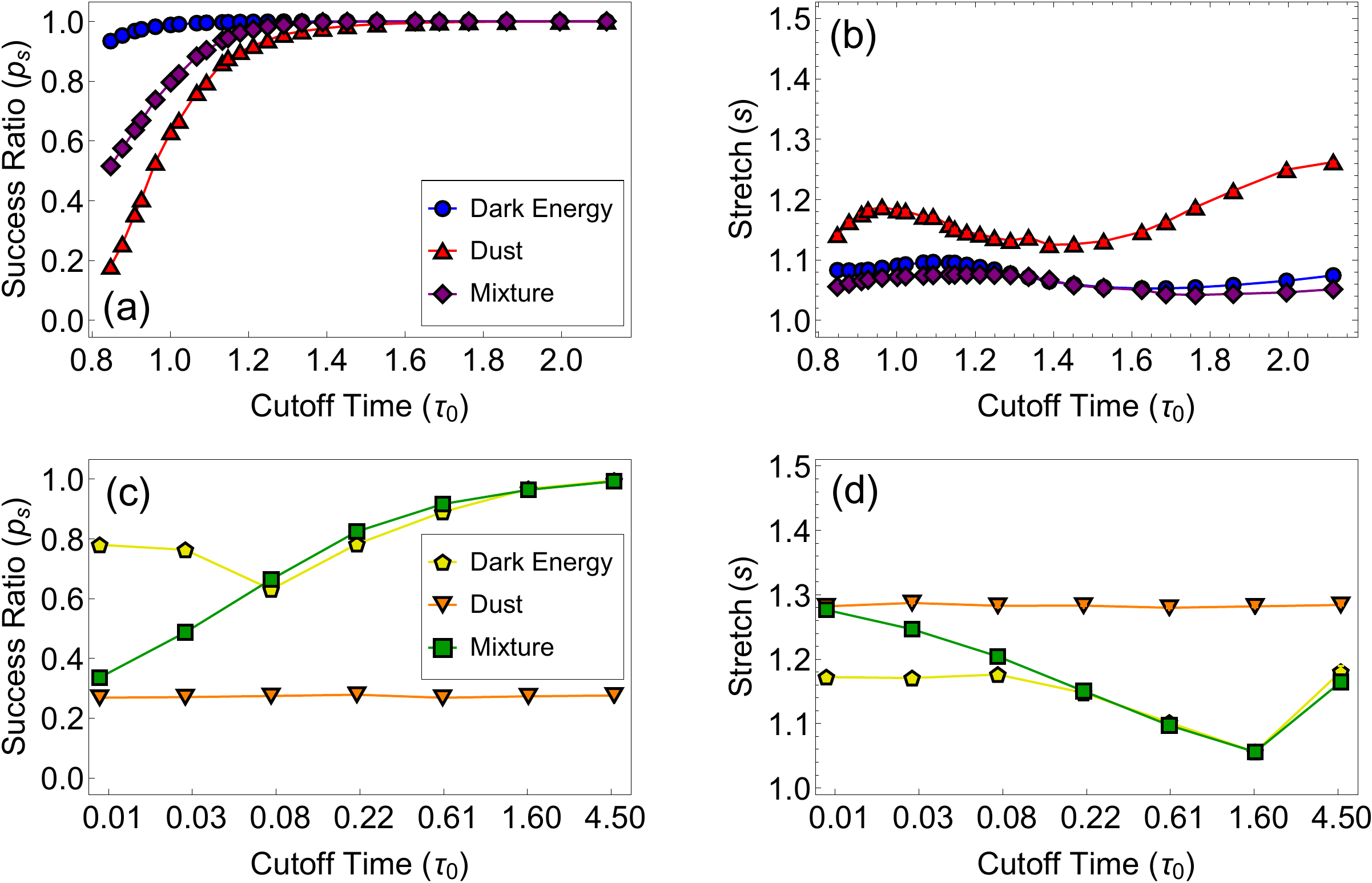}
\centering
\caption{\textbf{Navigability of random geometric graphs in the three manifolds.} In (a,b), corresponding to graphs in panels (a,b) in Fig.~\ref{fig3} where the sprinkling density and spatial cutoff are held constant $q=60$ and $\rho_0=6$, the success ratio increases toward $100\%$ as the temporal cutoff increases, while the average stretch remains low and close to $1$, especially for spacetimes with dark energy. In (c,d), the graph size and average degree are kept constant $N=2^{20}$ and $\bar{k}=10$ as described in the Methods. The success ratio and stretch in this case depend only on manifold geometry. The average stretch is still low, especially for the manifolds with dark energy. However, the success ratio increases to $100\%$ only for spacetimes with dark energy, while for the dust manifold it is a constant below $100\%$, which does not depend on the cutoff time.}
\label{fig5}
\end{figure}
\clearpage
Figures~\ref{fig5}(a,b) show that if the dimensionless sprinkling density $q$ is held constant as the temporal cutoff increases, the success ratio $p_s$ increases to $100\%$ in all the three manifolds, while the average stretch remains low and close to its minimum value~$1$, especially in the manifolds with dark energy. However, the average degree grows quickly with the temporal cutoff in this case, Eq.~\eqref{eq:degrees} and Figure~\ref{fig3}, and the success ratio and stretch depend both on the manifold geometry and on the average degree. Indeed, all other things equal, e.g., the same patch of the same manifold with the same spatial and temporal cutoff, the higher the average degree, the higher the navigability, i.e., the higher the success ratio and the lower the stretch, because the larger the number of neighbors that each node has, the higher the chances that the node has a neighbor that does not lead to a loop, and the higher the chances that the next-hop neighbor is closer to the geodesic to the destination in the manifold, thus minimizing the stretch. \par

To disentangle the dependency of navigability on manifold geometry from its dependency on the graph properties, the average degree and graph size, we select for different temporal cutoffs, different sprinkling densities and spatial cutoffs such that the average degree and graph size stay constant as the temporal cutoff increases, see the Methods. In this case, the navigability metrics depend only on the geometry of the manifold. \par

The results in Figure~\ref{fig5}(c,d) show that in this case, while the average stretch remains low, especially in the manifolds with dark energy, the success ratio depends strongly on the presence of dark energy in the spacetime. In spacetimes with dark energy, the success ratio still quickly reaches $100\%$, while in the dust-only spacetime, it is a constant below $100\%$, i.e., does not increase with time. \par

We thus conclude that unless dark energy is present, random graphs in Lorentzian geometries are not navigable as their success ratio is a constant below $100\%$, independent of the temporal cutoff. Only in spacetimes with dark energy and asymptotically de Sitter geometry, the success ratio quickly reaches its maximum value of $100\%$, so that such spacetimes, including the spacetime of our universe, are fully navigable with respect to all geodesically connected pairs of nodes. This result deserves a discussion.

\section{Discussion}
\begin{figure}[!b]
\includegraphics[width=0.5\textwidth]{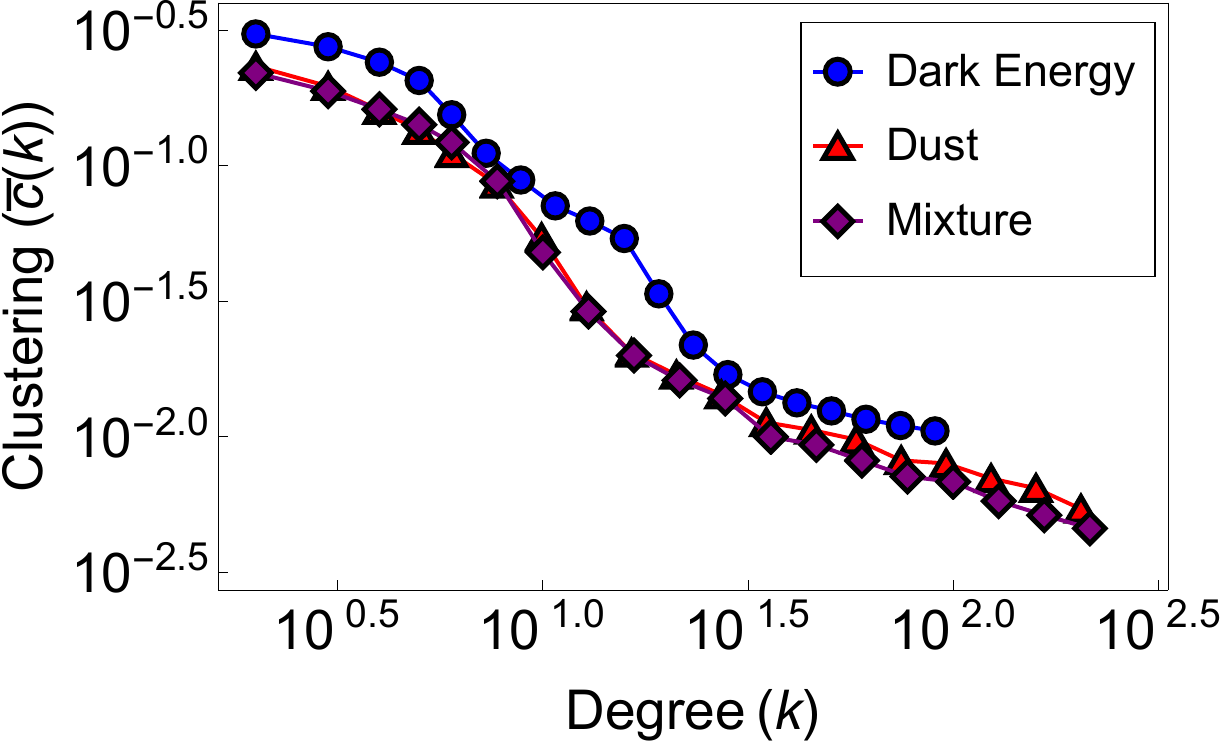}
\centering
\caption{\textbf{Clustering in Lorentzian RGGs.} The figure shows the average clustering $\bar{c}(k)$ of nodes of degree $k$ in random geometric graphs with $q=60,\rho_0=6,\tau_0=0.84$ in the three studied manifolds. The mean clustering excluding nodes with $k=\{0,1\}$ in the de Sitter, Einstein-de Sitter, and mixed manifolds are $\bar{c}_E=0.145$, $\bar{c}_D=0.164$, and $\bar{c}_M=0.166$, respectively.}
\label{figS3}
\end{figure}

\begin{figure}[!pt]
\includegraphics[width=\textwidth]{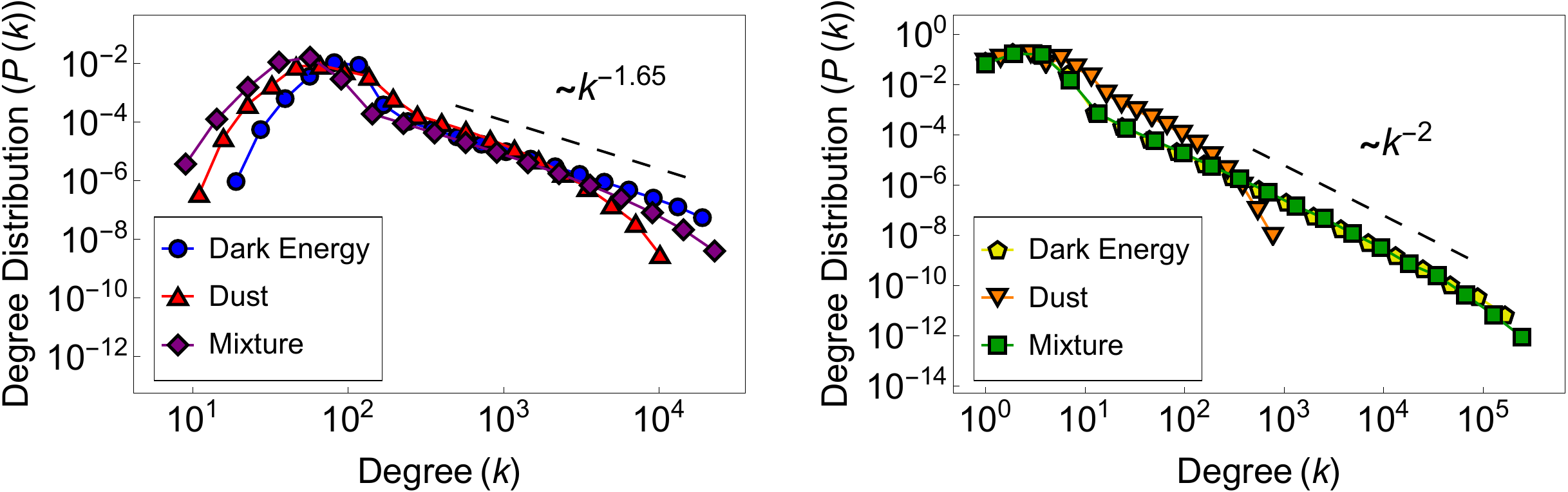}
\centering
\caption{\textbf{Degree distribution in Lorentzian RGGs.} Panels (a) and (b) show the degree distribution in the random geometric graphs in the three considered manifolds in the constant-$q$ and constant-$N,\bar{k}$ experiments, respectively, at the largest considered cut-off times $\tau_0$. Specifically, in panel (a) $q=60$, $\bar{k}=130$, $N=2518528$, $\tau_0=2.11$, and $\rho_0=6$, while in panel (b) $q=0.564$, $\bar{k}=10$, $N=2^{20}$, $\tau_0=4.64$, and $\rho_0=1.68$.}
\label{figS4}
\end{figure}

The navigability of random hyperbolic graphs and real networks is the higher, the lower the power-law degree distribution exponent $\gamma$, and the stronger the clustering~\cite{BoKrKc08,KrPa10}. Clustering in Lorentzian random geometric graphs considered here is not so strong, Figure~\ref{figS3}, primarily because of their higher dimensionality~\cite{Dall2002Random,Dhara2016Solvable} (3+1 versus 1+1) and small cut-off times, but the tails of the degree distributions, Figure~\ref{figS4}, in the graphs in the manifolds with dark energy follow power laws in the full agreement with the earlier results~\cite{KrKi12} showing that random geometric graphs in asymptotically de Sitter spacetimes have double power-law degree distributions with $\gamma=3/4$ at low degrees $k<q$ and $\gamma\to2$ at high degrees $k>q$. We note however that those results were derived only for the two limits of $\tau_0 \ll 1$ and $\tau_0 \gg 1$.\par

\begin{figure}[!pb]
\includegraphics[width=\textwidth]{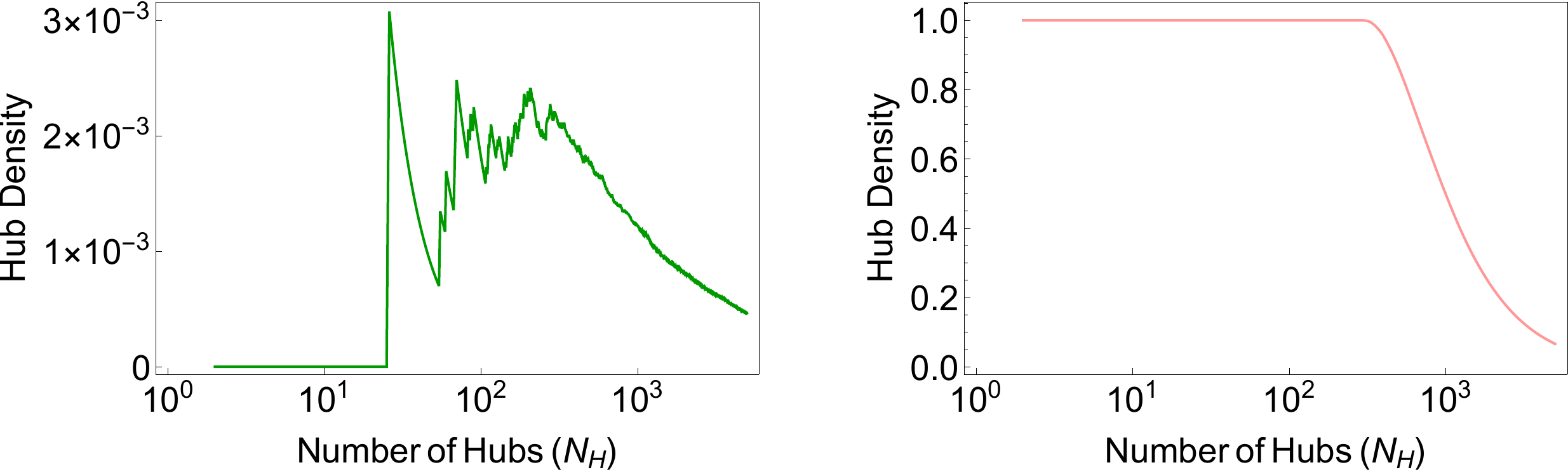}
\centering
\caption{\textbf{Hub density in Lorentzian and hyperbolic random graphs.} The hub density is defined as the number of links among the $N_H$ nodes with largest degrees, divided by the maximum possible number $N_H\choose2$ of such links. Panels~(a,b) compare the hub density in two random graphs of the same size $N=2^{20}$ and average degree $\bar{k}=10$. Panel~(a) shows the data for the mixed-content (M) Lorentzian manifold graph with $\rho_0=1.68$ and $\tau_0=4.64$, while panel~(b) shows the same data for the hyperbolic graph generated using \url{http://named-data.github.io/Hyperbolic-Graph-Generator/} with parameters $N=2^{20}$, $\bar{k}=10$, $\gamma=2$, and $T=0$ (the resulting radial cutoff is $\rho_0=32.36$). There are exactly zero links between 25 largest-degree nodes in the Lorentzian graph, while the subgraph induced by the first 103 highest-degree nodes in the hyperbolic graph is the complete graph.}
\label{figS5}
\end{figure}
More interestingly, as evident from Figure~\ref{fig1}, hubs, i.e., high-degree nodes, in random geometric graphs in Lorentzian manifolds are \emph{not} densely interconnected, Figure~\ref{figS5}, versus random hyperbolic graphs and real networks exhibiting strong rich club effects~\cite{Zhou2004RichClub,BoLaMoChHw06}. This hub disconnectedness is a characteristic feature of any Lorentzian random geometric graphs because nodes of similar degrees in them have similar time coordinates, and thus tend to be not connected since they do not lie within each other light cones with high probability. This observation may be puzzling, as it brings up the question of how Lorentzian graphs can be navigable at all, since one might intuitively think that geometric routing paths must go through the network core~\cite{BoKrKc08}, and if the hubs in this core are not all densely interconnected, then routing would fail with high probability. \par

\begin{figure}[!t]
\includegraphics[width=0.5\textwidth]{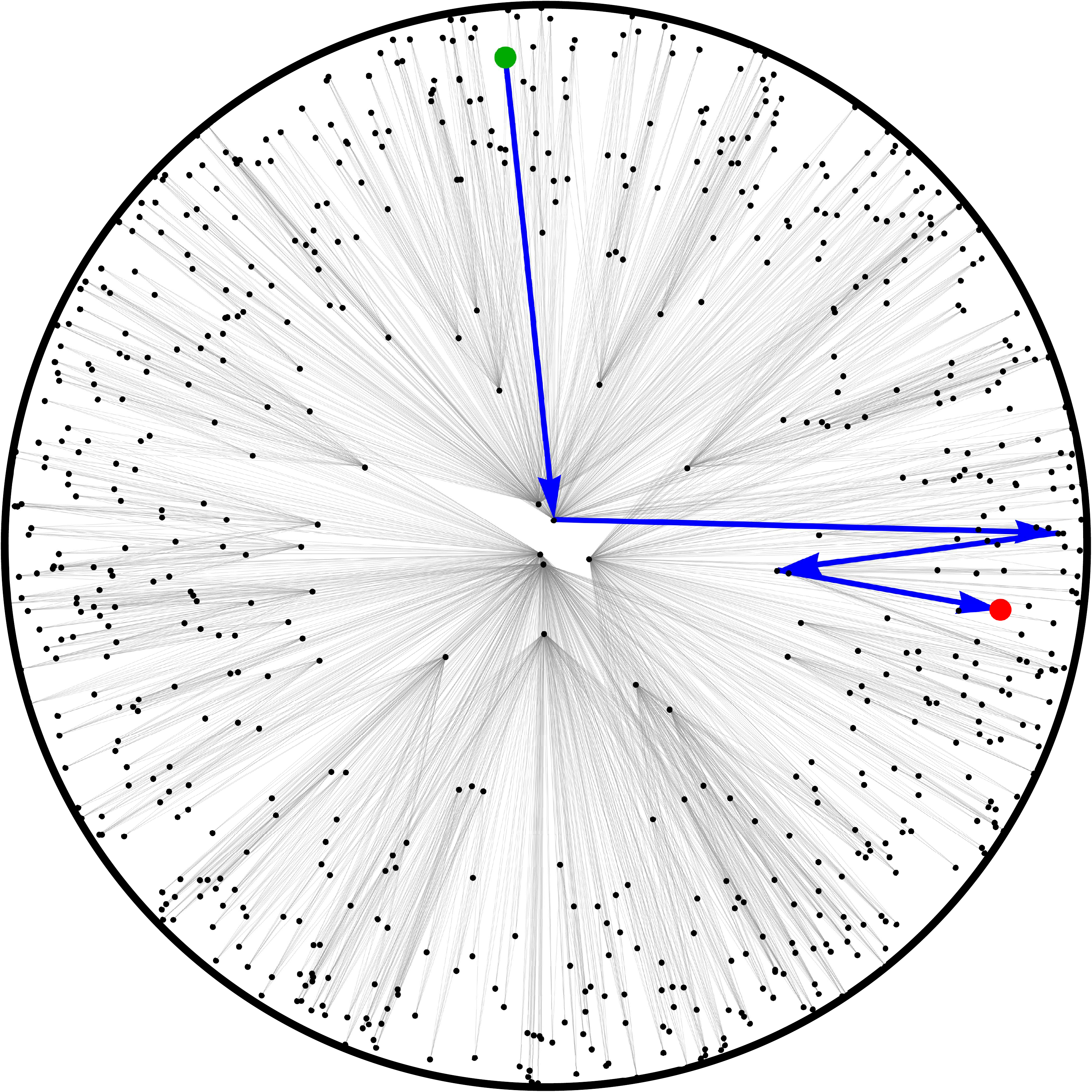}
\centering
\caption{\textbf{A typical navigation path in a Lorentzian RGG.} The figure shows the greedy geometric routing navigation path from the spacelike-separated green source and red destination in the same graph as in Figure 1~in the main text. The greedy path, which is also the shortest (stretch-$1$) path in the graph, alternates between hubs and peripheral nodes. Any timelike-separated pairs of nodes are directly linked, resulting in trivial one-hop stretch-$1$ paths.}
\label{figS6}
\end{figure}

This intuition turns out to be wrong, and the resolution of this puzzle lies in that the structure of geometric routing paths in Lorentzian graphs is completely different from the one in Riemannian graphs~\cite{BoKrKc08,KrPa10}. Specifically, the Lorentzian path structure exhibits a peculiar periphery-core zigzagging pattern illustrated in Figure~\ref{figS6}.
This pattern, in which subsequent hops tend to lie close to light cone boundaries, is caused by a completely different nature of Lorentzian geometry and the structure of geodesics in it, versus the Riemannian case, making the graphs navigable even though their cores are sparse.

As a final remark, this navigation pattern also shows that navigability of \emph{directed} causal sets based on random geometric graphs in Lorentzian manifolds is not so interesting. If links are directed in the past$\to$future time direction, then geometric routing respecting link direction and starting from a given source node succeeds only for destination nodes lying in the future light cone of the source. All such destinations are directly connected to the source. Navigation fails for any other source-destination pairs, including all spacelike-separated pairs of nodes, because paths between them necessarily involve hops in the future$\to$past direction.

\begin{figure*}[!t]
\includegraphics[width=\textwidth]{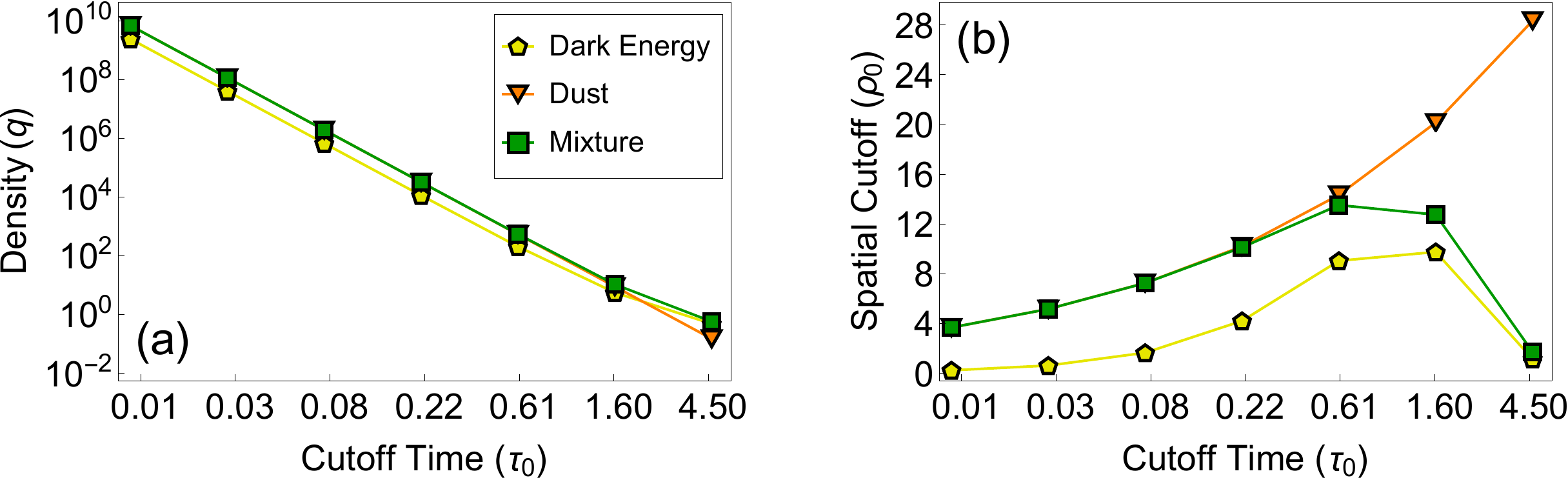}
\centering
\caption{\textbf{Rescaled sprinkling density and spatial cutoff as functions of the temporal cutoff in Fig.~\ref{fig5}(c,d).}}
\label{fig7}
\end{figure*}

\section{Methods}
{\bf Parameter range selection.} As discussed in the main text, the three parameters of the studied graph ensembles are the rescaled sprinkling density $q=\delta\lambda^4$, (rescaled) spatial cutoff $\rho_0=(\alpha/\lambda)r_0$ ($\rho_0=r_0$ in de Sitter spacetime), and rescaled cutoff time $\tau_0=t_0/\lambda$, which taken together determine the graph size $N$ and average degree $\bar{k}$ via (\ref{eq:nodes},\ref{eq:degrees}). In simulations, especially in navigability experiments, we have the following constraints: 1)~the graphs cannot be too large so that they fit into memory, $N\lesssim2^{21}$; 2)~the average degree cannot be too low so that the graphs are above the percolation threshold, $\bar{k}\gtrsim5$; 3)~the spatial cutoff must be sufficiently larger than the temporal cutoff, so that the spatial boundary effects are negligible and we can rely on~\eqref{eq:degrees}; 4)~we want to explore the most interesting region of $\tau_0\sim1$, corresponding to the rescaled dark energy density $\Omega_\Lambda$ changing over essentially an entire range of its values between $0$ and $1$. \par

In experiments with constant $q=60$, Figs.~\ref{fig3},\ref{fig5}(a,b), we select constant $\rho_0=6$ such that the average degree observed in simulations is within the error bound of $5\%$ from~\eqref{eq:degrees} for the largest considered value of $\tau_0>1$. This largest value of $\tau_0$ and the value of $q=60$ are determined in turn by the rest of the constraints above---decreasing $q$ would decrease the graph sizes, but would also decrease the average degree. The largest considered value of $\tau_0$ correspond to the largest graph sizes that fit into the memory, while the lowest value of $\tau_0$ is determined by the average degree value just above the percolation threshold. \par

In experiments with constant $\bar{k}=10$ and $N=2^{20}$, Fig.~\ref{fig5}(c,d), $q$ and $\rho_0$ as functions of $\tau_0$ are varied as solutions of the systems of equations~(\ref{eq:nodes},\ref{eq:degrees}), Fig.~\ref{fig7}. For all the considered values of temporal cutoff $\tau_0$, the spatial cutoff $\rho_0$ is sufficiently larger than $\tau_0$, so that the average degree is within the $5\%$ error bound from its theoretical fixed value $\bar{k}=10$, except for the largest value of $\tau_0=4.64$, where the average degrees in the de Sitter and mixed manifold cases are $8.13$ and $8.48$, respectively. \par

The non-monotonic dependency of the success ratio $p_s$ on the cutoff time $\tau_0$ in the dark energy manifold in Fig.~\ref{fig5}(c) is likely due to an interplay between increasing $\tau_0$, tending to increase $p_s$, and decreasing $q$, Fig.~\ref{fig7}(a), tending to decrease $p_s$, in the absence of spacetime singularity at $\tau_0$. The exact reason why this interplay is not important in the other two spacetimes that have this singularity is unclear. The non-monotonic behavior of stretch in Fig.~\ref{fig5}(b,d) is not surprising, since stretch is computed for successful paths only, whose percentages vary as shown in Fig.~\ref{fig5}(a,c). In particular, we have verified that the stretch increase in spacetimes with dark energy for the largest value of $\tau_0$ in Fig.~\ref{fig5}(d) is not due a below-the-borderline value of $\rho_0$: we have densely sampled the region of $\tau_0\in[1.6,4.5]$ (not shown), and found that the intermediate stretch values for these two manifolds lie on smooth curves connecting the two shown data points, while for most of these intermediate values of $\tau_0$, the value of $\rho_0$ is above the $5\%$ $\bar{k}$-accuracy borderline discussed above. \par

{\bf Statistics, simulations.} All the data shown in Figs.~\ref{fig3},\ref{fig5} are averaged over ten random graphs if $N<2^{20}$, over five graphs if $N=2^{20}$, or over three graphs if $N>2^{20}$. All the error bars in these figures are smaller than the symbol sizes. To generate graphs efficiently, we use OpenMP to generate node coordinates in parallel. Nodes are then linked using an NVIDIA K20m GPU via the CUDA library, since this step is the slowest when $N$ is large. While the linking algorithm is still $O(N^2)$, GPU parallelization offers a speedup of several orders of magnitude.

\section*{Acknowledgements}
We thank Alexander Vilenkin, Jaume Garriga, Ken Olum, Ji\v{r}\'{i} Podolsk\'{y}, David Rideout, and Michel Buck for useful discussions. This work was supported by NSF grants No.\ CNS-1442999 and CCF-1212778; and by ARO grant No.\ W911NF-16-1-0391.

\newpage
\bibliographystyle{naturemag}
\bibliography{paper,bib}

\end{document}